\documentclass[twocolumn]{emulateapj}
\usepackage{apjfonts}
 \usepackage{graphicx}
 \usepackage{amssymb}
 \usepackage{dsfont}
 \usepackage{esint}
 \usepackage{textcomp}
 
   \submitted{Accepted for publication in The Astrophysical Journal}
  
 \newcommand{\pf}{f_\perp}
 \newcommand{\po}{\theta_\perp}
 \newcommand{\pr}{R_\perp}

\begin{document}

\bibliographystyle{apj}

\title{Empirical Constraints on the Oblateness of an Exoplanet}

\author{Joshua A.~Carter \& Joshua N.~Winn}

\affil{Department of Physics, and Kavli Institute for
  Astrophysics and Space Research, \\Massachusetts Institute of
  Technology, Cambridge, MA 02139 \\
  carterja@mit.edu, jwinn@mit.edu}

\begin{abstract}

  We show that the gas giant exoplanet HD~189733b is less oblate than
  Saturn, based on {\em Spitzer Space Telescope} photometry of seven
  transits. The observable manifestations of oblateness would have
  been slight anomalies during the ingress and egress phases, as well
  as variations in the transit depth due to spin precession. Our
  nondetection of these effects gives the first empirical constraints
  on the shape of an exoplanet. The results are consistent with the
  theoretical expectation that the planetary rotation period and
  orbital period are synchronized, in which case the oblateness would
  be an order of magnitude smaller than our upper limits. Conversely,
  if HD~189733b is assumed to be in a synchronous, zero-obliquity
  state, then the data give an upper bound on the quadrupole moment of
  the planet ($J_2 < 0.068$ with $95\%$ confidence) that is too weak
  to constrain the interior structure of the planet. An Appendix
  describes a fast algorithm for computing the transit light curve of
  an oblate planet, which was necessary for our analysis.

\end{abstract}

\keywords{methods: numerical---stars: planetary systems---techniques:
  photometric}

\maketitle

\section{Introduction}

Planets are not exactly spherical. Departures from sphericity are
caused by rotation, external gravitational tides, and material
rigidity. For gas giant planets, material rigidity is not important,
but tides and rotation are very important. The tidal bulges raised on
``hot Jupiters'' by their parent stars allow angular momentum to be
exchanged between the planetary rotational and orbital motions, and
result in long-term energy dissipation (Goldreich \& Soter 1966, Peale
1999, Murray \& Dermott 2000). As for rotation, both Jupiter and
Saturn are visibly flattened due to centrifugal forces, with their
polar radii shorter than equatorial radii by 6.5\% and 9.8\%,
respectively (Lindal et al.\ 1981, 1985).

It would be of great interest to determine the shapes of planets
outside of the Solar System. Knowledge of the equilibrium shape of an
exoplanet would provide information on its rotation rate and internal
density structure, which in turn would give clues about its formation
and evolution. One way this knowledge might be gained is through
precise photometry of exoplanetary transits. Seager \& Hui~(2002) and
Barnes \& Fortney~(2003) calculated the difference between the transit
light curve of an oblate planet and of a spherical planet with the
same cross-sectional area. Unfortunately the oblateness-induced signal
is expected to be quite small for most of the currently known
transiting planets because their rotation periods are expected to be
tidally synchronized with their $\sim$3~day orbital periods (much
longer than the $\approx$10~hr rotation periods of Jupiter and
Saturn). For the representative case of HD~209458b, Barnes \&
Fortney~(2003) found the amplitude of the theoretically expected
oblateness-induced signal to be 0.1~$\mu$mag, well below the limiting
precision of any current photometer.

The goal of this work was to place the first {\it empirical}
constraints on the shape of an exoplanet.  While the assumption of
spin-orbit synchronization seems reasonable, it is worth checking on
such assumptions whenever possible. It is also useful to know the
upper limits that can be achieved with current data, with an eye
toward planning future observations and photometric instruments. By
selecting the most favorable planet and the best available data, we
find that an oblateness as large as that of Saturn can be ruled out,
and an oblateness as large as that of Jupiter is also ruled out if it
is accompanied by a moderate obliquity. In addition, we model the
effects of spin precession on transit light curves, which were not
considered by Seager \& Hui~(2002) or Barnes \& Fortney~(2003), and
provide a fast algorithm for calculating the light curve of an oblate
planet.

This paper is organized as follows. In \S~\ref{sec:physics} we review
the relevant physics and geometry. In \S~\ref{sec:spitred}, we present
seven {\em Spitzer} 8~$\mu$m transit observations of the well-studied
system HD~189733 (Bouchy et al.~2005). In \S~\ref{sec:analysis}, we
present the results of fitting a parameterized model to the data,
including the sky-projected oblateness and obliquity in addition to
the usual transit parameters. We consider the case of a fixed
orientation for the planet (\S~\ref{sec:sinprec}) as well as uniform
precession of the spin axis (\S~\ref{sec:conprec}). In the latter case
we are able to constrain the {\em true} oblateness and obliquity as
well as the precession period. In \S~\ref{sec:disc} we summarize our
methods and results and suggest some possible
extensions. Appendix~\ref{app:algorithm} describes the algorithm we
used to produce transit light curves of oblate planets.

\section{Review of Planetary Oblateness} \label{sec:physics}

\subsection{Rotation} \label{sec:rotation}

A uniformly rotating self-gravitating fluid takes on the figure of an
oblate spheroid, with its minimum diameter along the axis of rotation
(Eddington 1926). The shape may be quantified by the {\it oblateness}
(or {\it flattening}) parameter $f$, defined as
\begin{eqnarray}
f = \frac{R_{\rm eq}-R_{\rm pol}}{R_{\rm eq}} \label{eq:fdef}
\end{eqnarray}
where $R_{\rm eq}$ and $R_{\rm pol}$ are the equatorial and polar
radii, respectively (Murray \& Dermott 2000). The angle between the
polar axis and the orbital axis is the {\it obliquity} $\theta$.

The relationship between $f$ and the rotation period $P_{\rm rot}$ is
found by identifying the contours of constant potential, where the
potential has both gravitational and centrifugal terms. The
gravitational terms are
\begin{eqnarray}
V_g(r, \psi) =
-\frac{G M_p}{r} \left[ 1-\sum_{n=2}^{\infty} J_n \left(\frac{R_{\rm eq}}{r}\right)^n {\cal P}_2(\cos \psi)\right]
\label{eq:v-grav}
\end{eqnarray}
where $J_n$ are the spherical mass moments\footnote{$J_n =
  \frac{1}{M_p R^n_{\rm eq}} \int_0^{R_{\rm eq}} \int_{-1}^{+1} r^n
  {\cal P}_n(\mu) \rho(r,\mu) 2 \pi r^2~d\mu~dr$} associated with
rotation, $\psi$ is the planetary colatitude, and ${\cal P}_n$ is the
Legendre polynomial of degree $n$. The centrifugal terms are
\begin{eqnarray}
V_c(r,\psi) = \frac{1}{3} \Omega^2 r^2 \left[{\cal P}_2(\cos \psi)-1\right],
\label{eq:v-cent}
\end{eqnarray}
where $\Omega \equiv 2\pi/P_{\rm rot}$ is the rotational angular
frequency. Assuming that the quadrupole moment $J_2$ is the most important
moment, the total potential is
\begin{eqnarray}
V_{\rm tot}(r,\psi) =
&-&\frac{G M_p}{r} + 
\left( \frac{G M_p R_{\rm eq}^2}{r^3} J_2 + \frac{1}{3} \Omega^2 r^2\right)
{\cal P}_2(\cos \psi)\nonumber\\
 &-&\frac{1}{3} \Omega^2 r^2.
\label{eq:v-total}
\end{eqnarray}
Because the surface of the planet lies on an equipotential, $V_{\rm
  tot}(R_{\rm eq}, \frac{\pi}{2}) = V_{\rm tot}(R_{\rm pol},0)$,
giving
\begin{eqnarray}
-\frac{G M_p}{R_{\rm eq}} - 
\!\!\!\!\!\frac{1}{2}\left( \frac{G M_p}{R_{\rm eq}} J_2+ \Omega^2 R_{\rm eq}^2\right)
& = &\nonumber\\
&&\!\!\!\!\!\!\!\!\!\!\!\!\!\!\!\!\!\!\!\!\!\!\!\!\!\!\!\!\!\!\!\!\!\!\!\!-\frac{G M_p}{R_{\rm eq}} +
\frac{G M_p R_{\rm eq}^2}{R_{\rm pol}^3} J_2.
\end{eqnarray} 
This equality implies (to leading order in $f$)
\begin{eqnarray}
f & = & \frac{3}{2} J_2 + \frac{1}{2} \frac{\Omega^2 R_{\rm eq}^3}{G M_p}. \label{eq:f}
\end{eqnarray}
Derivations of this relation are also given by Murray \&
Dermott~(2000) and Hubbard~(1984). Solving for $P_{\rm rot} = 2 \pi /
\Omega$, we find
\begin{eqnarray}
P_{\rm rot} & = & 2 \pi \sqrt{\frac{R_{\rm eq}^3}{G M_p \left(2 f-3 J_2\right) }}.
\label{eq:rot}
\end{eqnarray}
This result reduces to Eqn.~(6) of Seager \& Hui (2002) if the $J_2$
term is neglected.

For Solar system planets, $f$ is measured from direct images and $J_2$
is measured by monitoring elliptical orbits of satellites whose orbits
precess in response to the aspherical gravitational field.
Table~\ref{tab:solarsystem} gives $f$ and $J_2$ for planets in our
Solar System, to help place our results for the exoplanet HD~189733b in
context. The maximum possible value for $f$ occurs at the rotational
breakup limit, when the outward centrifugal acceleration equals the
gravitational acceleration at the equator. With reference to
Eqns.~(\ref{eq:v-grav}-\ref{eq:v-cent}), this criterion gives $f<
\frac{1}{2}+\frac{3}{4} J_2$.

\begin{deluxetable}{lcl}
\tablecolumns{3}
\tablewidth{0pt}
\tablecaption{Shape parameters of Solar system planets}
\tablehead{\colhead{Planet} & \colhead{Oblateness $f$} & \colhead{$J_2$}}
\startdata
Mercury & 0.00012 & 0.000060 \nl
Venus & 0.00009 & 0.000004 \nl
Earth & 0.00350 & 0.001083 \nl
Mars & 0.00520 & 0.001960 \nl
Jupiter & 0.06487& 0.014736 \nl
Saturn & 0.09796 & 0.016298 \nl
Uranus & 0.02293& 0.003343 \nl
Neptune & 0.01708 & 0.003411
\enddata
\label{tab:solarsystem}
\tablerefs{Murray \& Dermott (2000), Barnes \& Fortney (2003), Hubbard (1984).}
\end{deluxetable}

Another relation between $J_2$ and $f$ can be obtained by assuming the
planet's interior to be in a state of hydrostatic equilibrium, leading
to the Darwin-Radau approximation,
\begin{eqnarray}
\frac{J_2}{f} \approx -\frac{3}{10} + \frac{5}{2} \mathds{C}-\frac{15}{8} \mathds{C}^2. \label{eq:dr}
\end{eqnarray}
Here, $\mathds{C}$ is the dimensionless moment of inertia, defined
such that $\mathds{C} M_p R_{\rm eq}^2$ is the moment of inertia about
the spin axis (Murray \& Dermott 2000). The moment of inertia of Solar
system planets is not directly measurable but models of gas giant
interiors suggest $\mathds{C} \approx 0.23$ (Hubbard \& Marley 1989).

\subsection{Sky projection} \label{sec:proj}

A transit by an oblate planet will produce a slightly different light
curve than a transit by a spherical planet of the same cross-sectional
area. The differences are most pronounced during the ingress and
egress phases of the transit. There are also slight differences during
the complete phase of the transit, due to the limb darkening of the
stellar photosphere. For illustrations of these effects we refer the
reader to Seager \& Hui (2002) and Barnes \& Fortney~(2003).

The shape of a single transit light curve will depend only on the {\it
  sky projection} of the planet's figure at the time of the transit.
The sky projection of an oblate or prolate spheroid (or any triaxial
ellipsoid) is an ellipse. We define the {\it projected oblateness}
$\pf$ as $(a-b)/a$, where $a$ and $b$ are the lengths of the major and
minor axes of the ellipse. We also define the {\it projected
  obliquity} $\po$ as the angle between the major axis and the transit
chord. The relations between these projected quantities and the
unprojected quantities $f$ and $\theta$ can be found with elementary
geometry:
\begin{eqnarray}
\pf & = & 1-\sqrt{\sin^2\theta'+\left(1-f\right)^2 \cos^2\theta'}, \label{eq:projob} \\
\tan\po &=& \tan\theta~\sin\phi, \label{eq:beta}
\label{eq:proj}
\end{eqnarray}
where $\theta'$ is the angle between the planetary rotation axis and
the sky plane, given by
\begin{eqnarray}
\cos^2\theta' & = & \sin^2\theta \sin^2\phi + \cos^2\theta, \label{eq:costhetap}
\end{eqnarray}
and $\phi$ is the azimuthal angle of the line of nodes between the
planetary equator and the orbital plane, with $\phi=0$ corresponding
to the case when the rotation axis is tipped toward the observer.  A derivation of
Eqn.~(\ref{eq:projob}) is also given by Barnes (2009).

The calculation of the transit light curve of an oblate planet is
computationally intensive, because the intersection points between a
circle and an ellipse are nonanalytic, and because the precision of
the calculation must be very high in order to isolate the small
oblateness-specific effects. In Appendix~\ref{app:algorithm} we
describe a fast algorithm that we developed for our study, building on
the previous work by Seager \& Hui~(2002) and Barnes \&
Fortney~(2003).

\subsection{Spin precession} \label{sec:prec}

External gravitational forces will cause a planet's spin axis to
precess.  In particular, the spin axis of a rotationally-induced,
oblate spheroidal planet will precess in response to the gravitational
torque of its host star, with a period
\begin{eqnarray}
P_{\rm prec} =
\frac{2}{3} \frac{P_{\rm orb}^2}{P_{\rm rot}} \frac{\mathds{C}}{J_2}
\frac{1}{\cos \theta} \label{eq:prec}
\end{eqnarray}
where $\theta$ is the planetary obliquity (Ward 1975). This expression
assumes that the orbit is fixed, and in particular that the orbital
angular momentum is much larger than the rotational angular momentum,
a good approximation even for close-in giant planets rotating as fast
as Jupiter or Saturn.\footnote{In reality the orbital and spin axes
  both precess about the total angular momentum vector. The nodal
  precession of the orbit would be detectable in principle through
  changes in the transit impact parameter. However, even if HD~189733b
  were spinning as fast as Jupiter, the orbital angular momentum would
  exceed the rotational angular momentum by a factor of 1000, and the
  resulting inclination variation would not be detectable in the
  current data.}

For Solar system planets, the precession periods are much longer than
1~yr. For example, Saturn completes one precession cycle in $\approx$
$7 \times 10^6$~yr (ignoring the effect of moons and rings; Ward \&
Hamilon 2004). However, because $P_{\rm prec}$ scales as $P_{\rm
  orb}^2$, sufficiently close-in planets would precess with periods
short enough to be directly observable. If Saturn's orbital period
were changed to $P_{\rm orb}=3$~days it would precess with $P_{\rm
  prec} \sim 0.6$~yr. In \S~\ref{sec:conprec}, we consider how spin
precession would be manifested in a collection of transit light
curves.

\subsection{Tidal deformation} \label{sec:oblatetides}

Hot Jupiters are stretched radially (along the radius vector of the
orbit) due to the gravitational tide from the nearby host star. Tidal
dissipation causes important long-term changes in the orbital
parameters, including spin-orbit synchronization and orbital
circularization. The characteristic timescale for synchronization is
\begin{eqnarray}
  \tau & \approx & \frac{4}{9} Q\,\mathds{C} \left( \frac{R_{\rm eq}^3}{G M_p} \right) \omega_0
  \left(\frac{M_p}{M_\star}\right)^2 \left( \frac{a}{R_{\rm eq}} \right)^6 \label{eq:synch}
\end{eqnarray}
where $\omega_0$ is the initial
angular rotation frequency, $M_p$ is the planet's mass, $M_\star$ is
the stellar mass, and $Q$ is the specific dissipation factor of the
tidal oscillator (Goldreich \& Soter 1966, Hubbard 1984, Guillot et
al.\ 1996, Murray \& Dermott 2000). Many investigators adopt $Q\sim
10^{\rm 5-6}$ for gas giant planets, although there is little
empirical information about $Q$ and some indications that $Q$ may be
larger (see, e.g., Hellier et al.~2009).

The obliquity is also expected to be driven to zero as a result of
tidal evolution, although in an interesting and nonuniform manner
(Peale 1999). It is possible to maintain a nonzero obliquity in a
so-called Cassini state (Ward 1975, Winn \& Holman 2005) but for hot
Jupiters the high-obliquity states are unstable (Fabrycky et al.~2007,
Levrard et al.~2007).

It would be difficult to detect a radially-directed tidal bulge
through transit photometry. For tidally synchronized planets, the
amplitude of the rotational deformation is roughly $1/3$ that of the
tidal component (Murray \& Dermott 2000, pg.\ 156). However, the
long axis is closely aligned with the line of sight during transits,
and consequently the sky projection of the tidal bulge is smaller than
the bulge itself by a factor $a/R_\star$. This causes a reduction in
the transit signal of the radial bulge by an order of magnitude or
more, compared to the rotational bulge. Therefore in this paper we
consider only the deformation associated with rotation. Ragozzine \&
Wolf (2009) describe some potentially observable consequences of the
radial bulge.

\subsection{Expectations for HD 189733\lowercase{b}}

In this section we use the preceding formalism and the algorithm
described in the Appendix to compute theoretical light curves for the
transiting exoplanet HD~189733b. This particular exoplanet was chosen
because it is the most favorable case currently known for the
detection of oblateness-induced signatures, owing to the bright parent
star, large transit depth, midrange transit impact parameter, and the
availability of a large corpus of high-precision transit data. We took
the system parameters from Torres et al.~(2008) and considered two
different cases for the oblateness and obliquity.

\begin{figure*}[htbp] %  figure placement: here, top, bottom, or page
   \centering
   \plotone{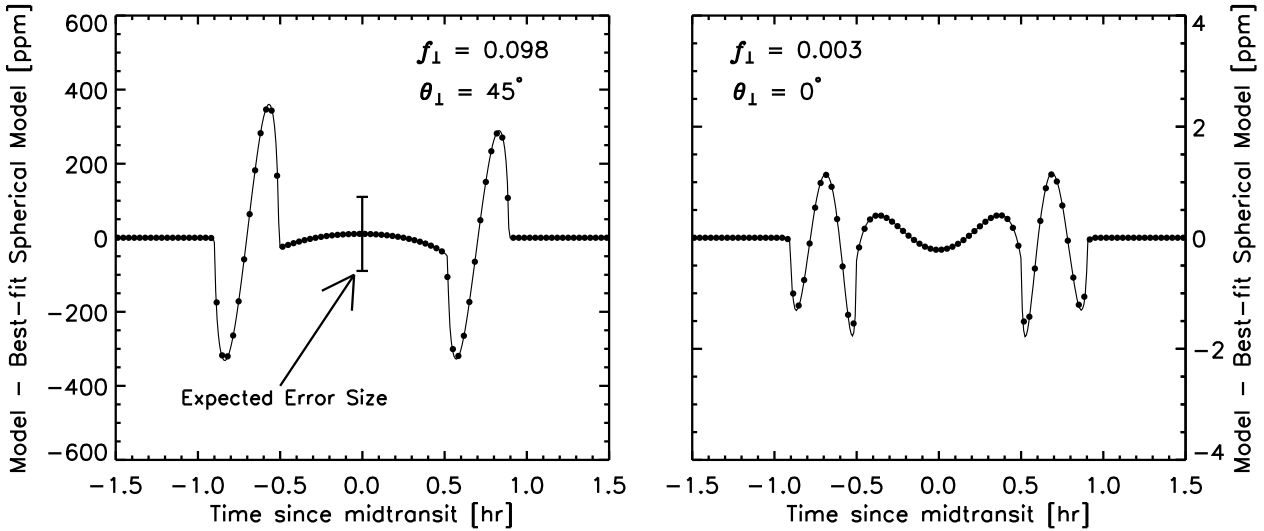}
   \caption{The theoretical oblateness-induced signal in the transit
     light curve of HD~189733b. Each panel shows the difference in the
     time series between the theoretical transit light curve of an
     oblate planet, and the best-fitting light curve of a spherical
     planet.  In computing the theoretical light curves, all the
     system parameters except oblateness and obliquity were taken from
     Torres et al.~(2008). {\it Left.}---Assuming HD~189733b is as
     oblate as Saturn and is more oblique ($\pf = 0.098$, $\po =
     45^\circ$).  The error bar indicates the expected 1$\sigma$ noise
     level of a composite light curve built from seven {\em Spitzer}
     observations, binned into 2~min intervals. {\it
       Right.}---Assuming $\pf=0.003$ and $\po=0$, as appropriate for
     the case of a tidally synchronized planet.}
   \label{fig:expectation}
\end{figure*}

First, we imagined that HD~189733b is as oblate as Saturn ($\pf =
0.098$), with a projected obliquity of $\po = 45^\circ$. To isolate
the signal due to oblateness, we calculated the theoretical transit
light curve for the oblate planet, and then subtracted the
best-fitting model for a spherical planet, following Barnes \&
Fortney~(2003). The residuals are shown in the left panel of
Figure~\ref{fig:expectation}. An error bar is also shown, representing
the forecasted precision of a 2~min sample based on the {\it Spitzer}
data that are currently available for this planet (see \S~3). This
calculation suggested that the {\it Spitzer} data would be capable of
placing physically meaningful constraints on the oblateness, and
motivated our further study.

Second, we assumed that HD~189733b has been tidally synchronized with
$P_{\rm rot} = P_{\rm orb} = 2.2$~days. This is reasonable because for
$Q=10^5$ and $\omega_{\rm prim} = 1.7 \times 10^{-4}$ s$^{-1}$
(Jupiter's estimated values; Guillot et al.~1996) we find $\tau \sim
10^6$~yr, much shorter than the estimated few-Gyr main-sequence age of
the star (Torres et al.~2008). Furthermore, the observation that the
orbit of HD~189733b is nearly circular (Winn et al.~2007b, Knutson et
al.~2007a) is independent evidence for tidal evolution, and the
theoretical synchronization timescale is shorter than the
circularization timescale. We estimated the rotationally-induced
oblateness using the Darwin-Radau relation and Eqn.~(\ref{eq:f}),
finding $f \approx 0.003$.  The right panel of
Figure~\ref{fig:expectation} shows the difference between a light
curve of an oblate planet with $\pf = 0.003$ and $\po=0$, and the
best-fitting model for a spherical planet.  The peak-to-peak amplitude
of the residuals is approximately $2\times10^{-6}$, below the
precision of the available data. Thus, if spin-orbit synchronization
has been achieved, we would expect a null result from an analysis of
the {\it Spitzer} data.

\section{Observations and data reduction}  \label{sec:spitred}

Transits of HD~189733b have been observed with many instruments and in
many bandpasses (see, e.g., Bouchy et al.~2005, Winn et al.~2007b, Pont
et al.~2007, Beaulieu et al.~2008, Miller-Ricci et al.~2008, D\'esert
et al.~2009). For our study the best available data are the 7
different transits that were observed with the {\it Spitzer Space
  Telescope} at a wavelength of 8~$\mu$m, with the InfraRed Array
Camera (IRAC; Fazio et al 2004). They form a homogeneous and precise
data set, and by virtue of the relatively long observing wavelength
the data are less affected by stellar variability, star spots and limb
darkening, which would otherwise confound the attempt to detect or
constrain oblateness. The first of the 7 IRAC time series was
presented by Knutson et al.~(2007a), and the other 6 time series are
based on observations by Agol et al.~(2009). The elapsed time between
the first and seventh transits was 1.6~yr or 268 transits. Assigning
epoch zero to the second transit (on UT~30~June~2007) the observed
transits have epoch numbers $-110$, 0, 1, 51, 62, 157, and 158.

All 7 transits were observed with the 8~$\mu$m channel of the IRAC
instrument in subarray mode, in which only a 32$\times$32 pixel
subarray of the detector is recorded. Images are obtained with a
cadence of 0.40~s (with an integration time of 0.32~s). The data are
packaged into post-calibration FITS files, each consisting of 64
images (representing a total of $26.5$~s of integration
time). Approximately 500 files span the transit of HD~189733b. In
addition to providing excellent time resolution, the subarray mode
avoids saturation even for stars as bright as HD~189733 ($V = 7.7$,
H{\o}g et al.~2000).

We began our data reduction with the post-calibration images,
downloaded from the {\em Spitzer} data archive. First, we inspected
each block of 64 images and determined the centroid of the target star
in each image. We excluded any image in which the centroid differed by
more than 0.05 pixels from the median position of all 64 images. Fewer
than 1\% of the images were excluded by this criterion, except for
epochs 1 and 157 in which $8\%$ and $5\%$ of images were excluded,
respectively. Next, we formed a mean image based on each block of 64
images, disregarding any pixels whose values deviated by more than
3.5$\sigma$ from the median value for that pixel. Fewer than 1\% of
the pixels were excluded by this criterion.

We then performed aperture photometry on the mean images, using a
circular aperture of radius 4.5 pixels centered on HD~189733.  To
estimate the background level, we also summed the flux within several
rectangular apertures located far away from both HD~189733 and its
fainter companion. The background level was subtracted from the
aperture sum. At this point, we had a time sequence of measurements of
the relative flux density of HD~189733.

Among users of the IRAC 8~$\mu$m channel there is a well-known
systematic effect which has been called the ``ramp,'' because it is
manifested as a gradual rise in the count rate at the beginning of an
observation. It is attributed to charge trapping in the detector. It
is generally modeled as a multiplicative, time-variable correction
(see, e.g., Knutson et al.~2007b, Gillon et al.~2007, Nutzman et
al.~2008). For each data set, we clipped the most strongly-varying
portion of the ramp, and modeled the rest as a quadratic function of
time (see \S~\ref{sec:sinprec}). We also clipped the long post-egress
portion of the time series by Knutson et al.~(2007a). Figure
\ref{fig:lcsfixed} shows the final time series, after correcting for
the ramp.

We assessed the noise characteristics of each time series with the
{\tt solveredwv} IDL routine\footnote{{\tt
    http://www.mit.edu/$\sim$carterja/code/} }. This algorithm fits
time-series data with a model in which the noise is an additive
combination of white noise and $1/f$ noise. The amplitudes of each
noise component are estimated from the data as described by Carter \&
Winn (2009). Many of our reduction parameters (aperture size,
thresholds for clipping, etc.)~were chosen by attempting to minimize
the time-correlated component of the noise. Our final time series had
white-noise amplitudes of 586, 571, 642, 544, 560, 607 and 536 ppm,
for epochs $-110$, 0, 1, 51, 62, 157, and 158, respectively.  In all
cases the amplitude of the $1/f$ component was smaller than 2~ppm.
The theoretical limiting precision due to photon-counting noise was
approximately 460~ppm.

\section{Analysis}  \label{sec:analysis}

\subsection{Fixed orientation} \label{sec:sinprec}

In the analysis described in this section, we assumed the orientation
of HD 189733b to be fixed in space over the 1.6~yr span of the
observations. The projected oblateness parameters $\pf$ and $\po$ were
taken to be constants. In the next section we will describe an analysis
in which that restriction was lifted.

The first step was a preliminary fit to all of the data, with the goal
of determining the midtransit time, out-of-transit flux, and the
coefficients of the ramp correction function for each epoch. These
quantities are weakly correlated with the other parameters describing
the light curves (including the projected oblateness and obliquity),
and thus may be fixed in the subsequent analysis without significantly
affecting the results.

The ramp correction function was
\begin{eqnarray}
C_{\rm ramp}(t; c_0, c_1; t_0) = 1+c_0 (t-t_0)+ c_1 (t-t_0)^2 \label{eq:ramp}
\end{eqnarray}
where $c_0$ and $c_1$ are adjustable parameters, and $t_0$ is a
particular time near midtransit. In addition to the ramp correction,
each light curve was described by an out-of-transit flux level $F_0$
and midtransit time $t_0$. There were also 7 parameters common to all
the light curves: the mean projected radius ratio $\pr/R_\star$ (where
$\pr= R_{\rm eq} \sqrt{1-\pf}$), orbital inclination $i$, normalized
orbital distance $a/R_\star$, quadratic limb-darkening coefficients
$u_1$ and $u_2$, projected oblateness $\pf$, and projected obliquity
$\po$. The limb darkening coefficients parameterize the stellar
brightness profile, $I_\star(r; u_1, u_2)$, defined as
\begin{eqnarray}
I_\star(\mu; u_1, u_2) =  I_\star(1) \left[1- u_1\left(1-\mu\right)-u_2 \left( 1-\mu\right)^2 \right] 
\end{eqnarray}
where $\mu = \sqrt{1-r^2}$ and $r$ is the sky-projected distance from the center of the star.

\begin{figure*}[htbp] %  figure placement: here, top, bottom, or page
   \centering
   \epsscale{1.05}
   \plotone{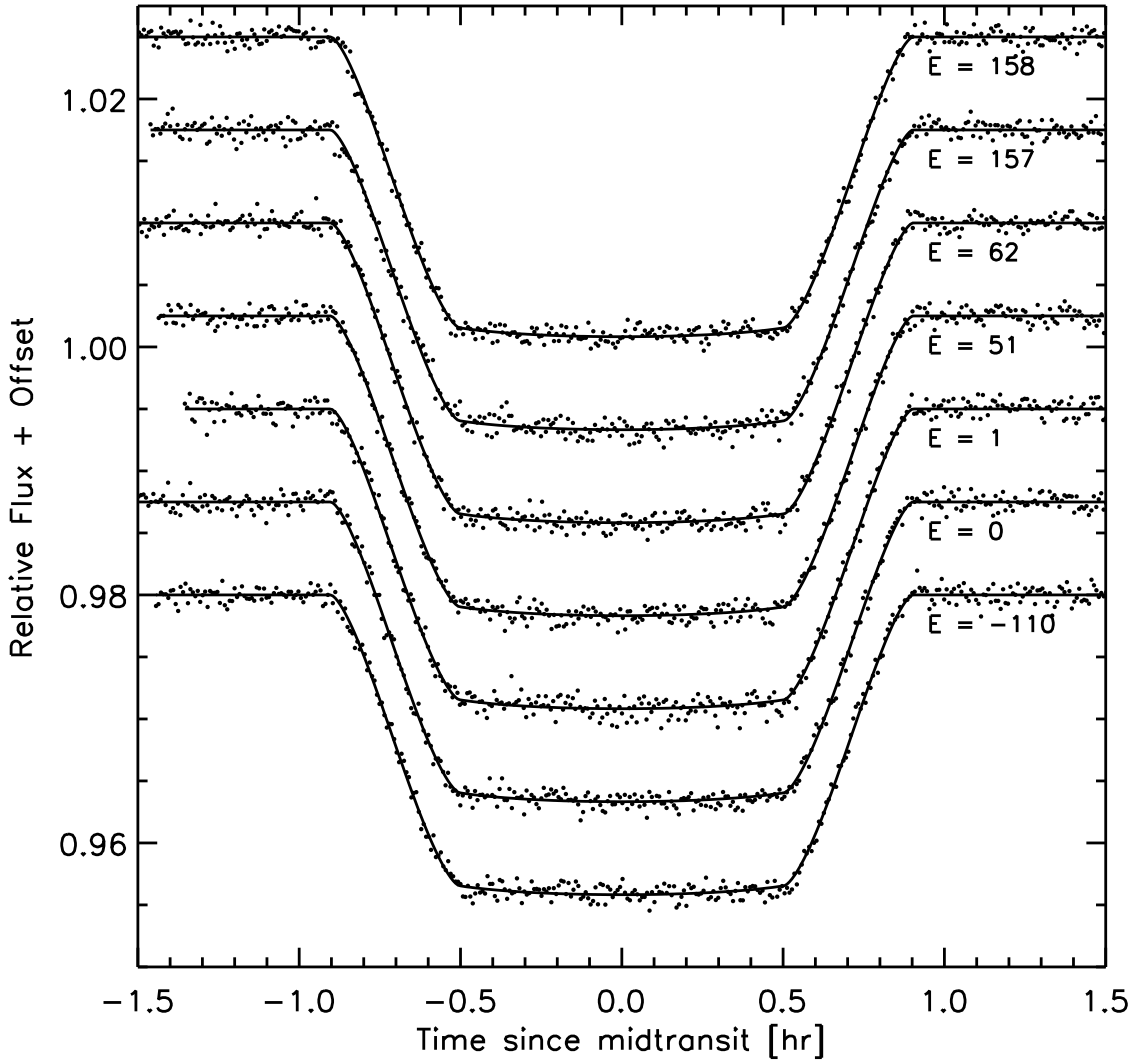} 
   \caption[Systematic corrected transit light curve data of seven
   {\em Spitzer} observations of HD~189733b.]{The relative 8~$\mu$m
     brightness of HD~189733 during 7 different transits of its giant
     planet, as measured with {\it Spitzer}/IRAC. The time series were
     corrected for the detector ``ramp,'' and offset in flux for
     display purposes. The solid curve is the best-fitting model. See
     \S~\ref{sec:sinprec} for details.}
   \label{fig:lcsfixed}
\end{figure*}

To derive the best-fitting parameter values we minimized the standard
$\chi^2$ statistic,
\begin{eqnarray}
\chi^2 = \sum_{o=1}^7 \sum_{i =0}^{N_o} \left[ \frac{F^o_i({\rm obs})-F^o_i({\rm calc})}{\sigma^o}\right]^2 \label{eq:chi2global}
\end{eqnarray}
where $o$ indexes the observation number, $N_o$ is the number of data
points in observation $o$, $F^o_i({\rm obs})$ is a measurement of the
relative flux of HD 189733 during observation $o$, $F^o_i({\rm calc})$
is the calculated flux at that time according to the model, and
$\sigma^o$ is the uncertainty in flux measurement during observation
$o$. We took $\sigma^o$ to be the white noise amplitude of each time
series as specified in \S~\ref{sec:spitred}. As described above, we
found the correlated noise component to be negligible. We minimized
$\chi^2$ using the AMOEBA routine (Press et al.~2007).  The
ramp-corrected light curves and the best-fitting model light curves
are shown in Figure~\ref{fig:lcsfixed}.

Then, after fixing the midtransit times, ramp correction parameters,
and out-of-transit flux levels at the best-fitting values, we
determined the credible intervals for the other parameters using a
Markov Chain Monte Carlo (MCMC) algorithm.\footnote{For background on
  the MCMC method, see Gregory (2005), and for examples of
  applications to transit light curves, see Holman et al.~(2006), Winn
  et al.~(2007a), or Burke et al.~(2007).} We found that the quadratic
limb darkening parameter was very poorly constrained by the data, and
therefore for subsequent work we fixed that parameter at the value
obtained in the preliminary fit. The parameters that were allowed to
vary in the chain were $\pr/R_\star$, $i$, $a/R_\star$, $u_1$, $T_0$,
$\pf$, and $\po$. Here, $T_0$ represents an overall timing offset of
all 7 transits, which is needed because the oblateness parameter is
correlated with an overall time shift.
  
In the MCMC, the jump-transition probability was proportional to
$\exp(-\chi^2/2)$ with $\chi^2$ given in Eqn.~(\ref{eq:chi2global}).
We used Gibbs sampling in the construction of a chain of $5\times10^5$
links.  We selected individual parameter jump sizes such that the
fraction of jumps accepted by a Metropolis-Hasting condition was
between $30\%$ and $60\%$ for each parameter. Uniform priors were used
for all parameters. We verified that the resulting posterior
distributions had converged sufficiently.

\begin{figure*}[htbp] % figure placement: here, top, bottom, or page
   \centering
   \epsscale{0.85}
   \plotone{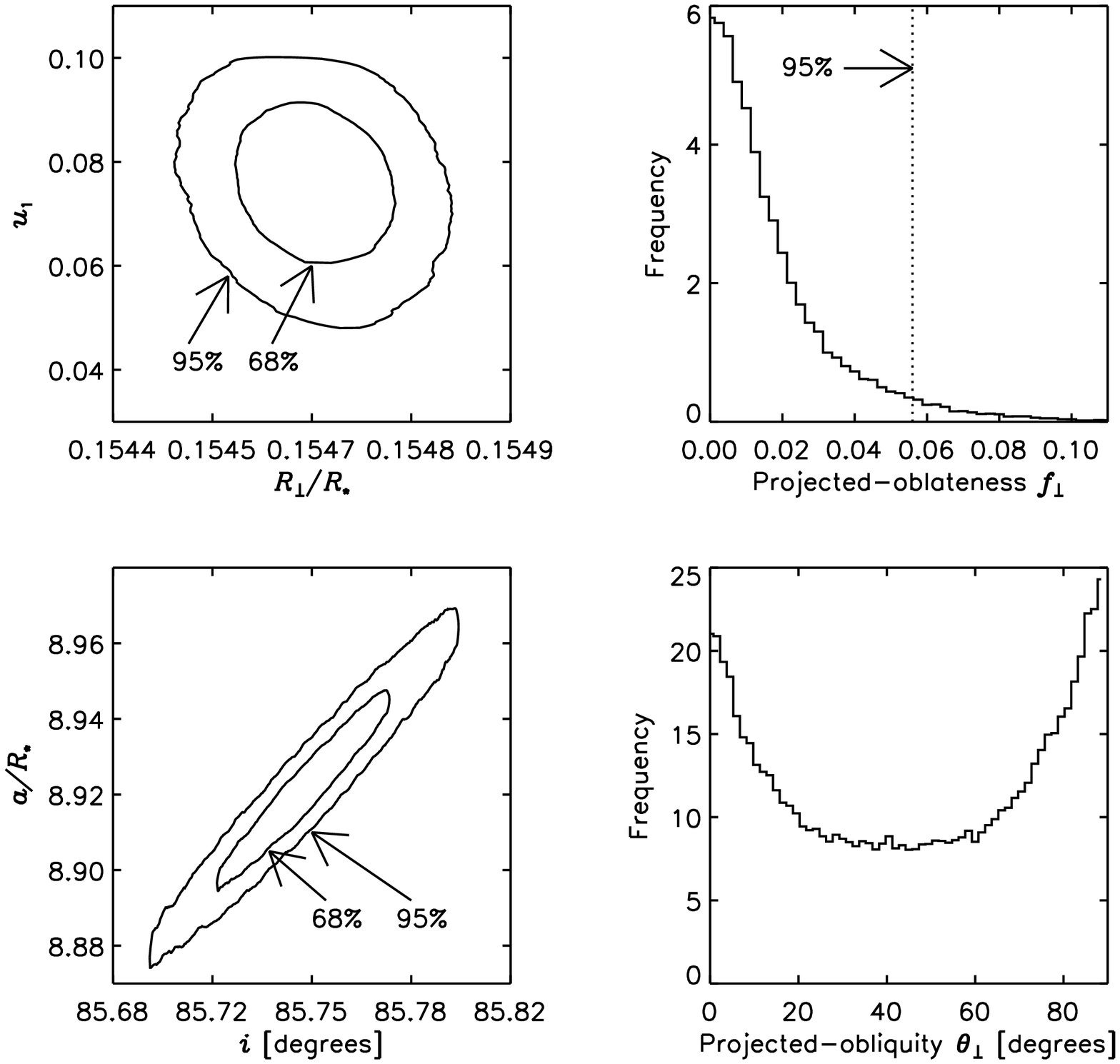}
   \caption{Joint confidence regions and posterior distributions for
     transit parameters and the projected shape parameters of
     HD~189733b, based on {\it Spitzer} observations of 7
     transits. The shape parameters were assumed to be
     time-independent.}
   \label{fig:joints}
\end{figure*}

\begin{figure*}[htbp] %  figure placement: here, top, bottom, or page
   \centering
   \epsscale{1.0}
   \plotone{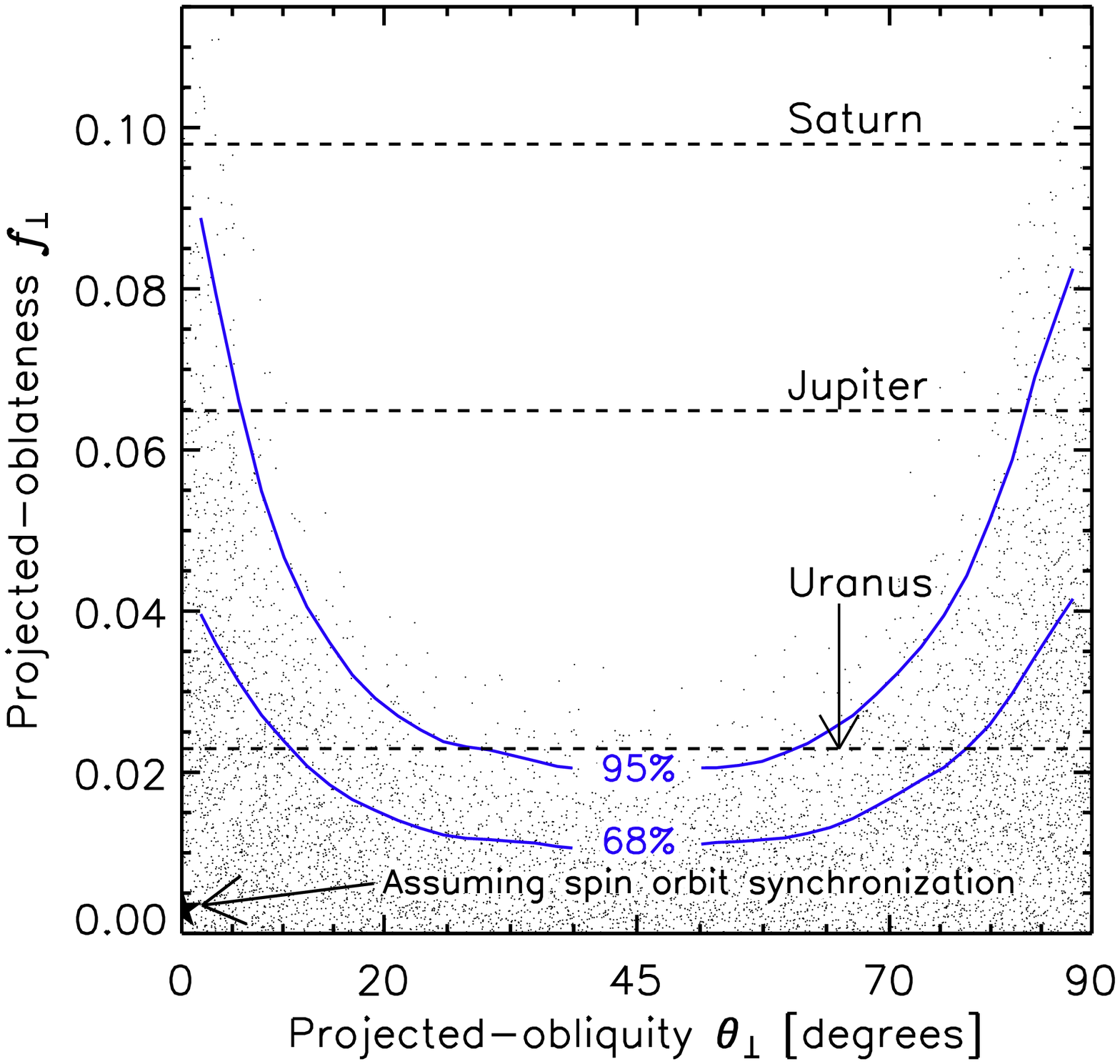}
   \caption{Posterior probability distribution for the projected
       oblateness and obliquity of HD~189733b, based on {\it Spitzer}
       observations of 7 transits. The shape parameters were assumed
       to be time-independent. The solid blue curves bound the regions
       containing $68\%$ or $95\%$ of the probability in the
       ($\pf$--$\po$) plane, marginalized over all other
       parameters. The black points are 10,000 representative values
       from the Markov chain, to illustrate the probability density.
       The star shows the expected shape parameters if HD~189733b is
       spin-orbit synchronized.  The dashed lines indicate the
       oblateness of Jupiter, Saturn, and Uranus, for comparison. }
   \label{fig:constraintsallob}
\end{figure*}

Table~\ref{tab:results1} gives the results. For most parameters we
report the median value of the posterior probability distribution,
along with error bars defined by the $15.85\%$ and $84.15\%$ levels of
the cumulative distribution. For the projected oblateness, we report
the $95\%$-confidence upper limit. The projected obliquity was
unconstrained. Figure~(\ref{fig:joints}) shows some of the posterior
probability distributions. Figure~\ref{fig:constraintsallob} shows the
constraints in the $\pf$--$\po$ plane. Only the results for $\po \ge
0$ are shown, as the constraints were found to be symmetric about $\po
= 0$. For comparison, we have indicated in the figure the oblatenesses
of Saturn, Jupiter and Uranus.

\begin{deluxetable*}{lll}
\tablecolumns{3}
\tablewidth{0pt}
\tablecaption{Model Results: Static Shape Parameters}
\tablehead{\colhead{Parameter} & Median & Uncertainty}
\startdata \nl
{\em $\!\!\!\!\!\!$Transit parameters:} &&\nl
$\pr/R_\star$ & 0.154679 & $\pm$0.000067 \nl
Orbital Inclination, $i$ [degrees] & 85.749 & $\pm$0.026 \nl
$a/R_\star$ & 8.924 & $\pm$0.022 \nl
Projected oblateness $\pf$ & 0 & $< 0.056$ ($95\%$ conf.)\nl
Projected obliquity $\po$ &unconstrained & \nl \nl
Limb darkening parameter $u_1$ & 0.076& $\pm$ 0.011\nl
Limb darkening parameter $u_2$ & 0.034 (fixed) & -- \nl
Midtransit time shift $T_0$ [s] & 0.0 & $\pm$1.3 \nl \nl
\tableline \nl
{\em $\!\!\!\!\!\!$Derived parameters$^{a}$:}&&\nl
Rotational Period [days]$^{b,c}$ & -- & $> 0.39$  ($95\%$ conf.) \nl
$J_2$$^d$ & -- & $< 6.8\times10^{-2}$ ($95\%$ conf.) \nl
\enddata
\tablecomments{(a) Assuming zero obliquity. (b) Calculated using
  Eqn.~(\ref{eq:rot}) with $R_\star = 0.756$ $R_\sun$, $M_p = 1.144$
  $M_{\rm Jup}$ (Torres et al.~2008). (c) Assuming the validity of the
  Darwin-Radau approximation (Eqn.~\ref{eq:dr}) with $\mathds{C} =
  0.225$.  (d) Calculated using Eqn.~(\ref{eq:f}) assuming $P_{\rm
    rot} = P_{\rm orb} = 2.218573$ days.  }
\label{tab:results1}
\end{deluxetable*}

As anticipated from the theoretical work of Seager \& Hui~(2002) and
Barnes \& Fortney~(2003), the projected oblateness is most tightly
constrained when the projected obliquity is near $45^\circ$. We
determined the $68\%$ and $95\%$ confidence regions in the
$\pf$--$\po$ plane with the following technique. First, we divided the
projected obliquity range into bins of constant width. Next, for the
chain links in each bin, we sorted the values of $\pf$ and determined
$\pf'$ such that $68\%$ ($95\%$) of the values were less than $\pf'$.
Fig.~(\ref{fig:constraintsallob}) shows the resulting confidence
curves.

Our analysis rules out a projected oblateness that is equal to
Saturn's oblateness with $>$95\% confidence, regardless of the
projected obliquity. A projected oblateness equal to Jupiter's
oblateness is also ruled out with $>$95\% confidence, except for
obliquities within about $7^\circ$ of either $0^\circ$ or
$90^\circ$. For a projected obliquity near $45^\circ$, we may also
exclude a projected oblateness equal to the oblateness of Uranus, with
95\% confidence.  Our constraints are consistent with the theoretical
expectation that HD~189733b is rotating in synchrony with its orbit
($\pf \approx f \approx 3\times10^{-3}$ and $\po \approx
0^\circ$).

The nondetection of oblateness suggests that the planet cannot be
rotating too quickly, but the results cannot be translated directly
into a lower bound on $P_{\rm rot}$ because of the sky projection.
Even a very rapidly rotating (and very oblate) planet is consistent
with the data as long as the planet's rotation axis is pointing at the
observer ($\theta = 90^\circ$ and $\phi=0$), leading to a circular
projected figure. If we assume $\theta\approx 0$ then we may set an
upper bound on $P_{\rm rot}$. We further assume that $J_2$ is given by
the Darwin-Radau relation (Eqn.~\ref{eq:dr}), with $\mathds{C} =
0.225$. Under these assumptions $P_{\rm rot} > 0.39$~d (9.4~hr) with
95\% confidence. Larger values of $\mathds{C}$ would correspond to
longer rotational periods. The median value of the posterior
distribution is $P_{\rm rot}$ is 0.65~days, but we do not attribute
any significance to that result, as the shape of the posterior
distribution (including the median) is strongly affected by our
assumption of a uniform prior in $\pf$. The lower bound on $P_{\rm
  rot}$ is less sensitive to the prior.

\begin{figure}[htbp] %  figure placement: here, top, bottom, or page
\centering
\epsscale{1.1}
\plotone{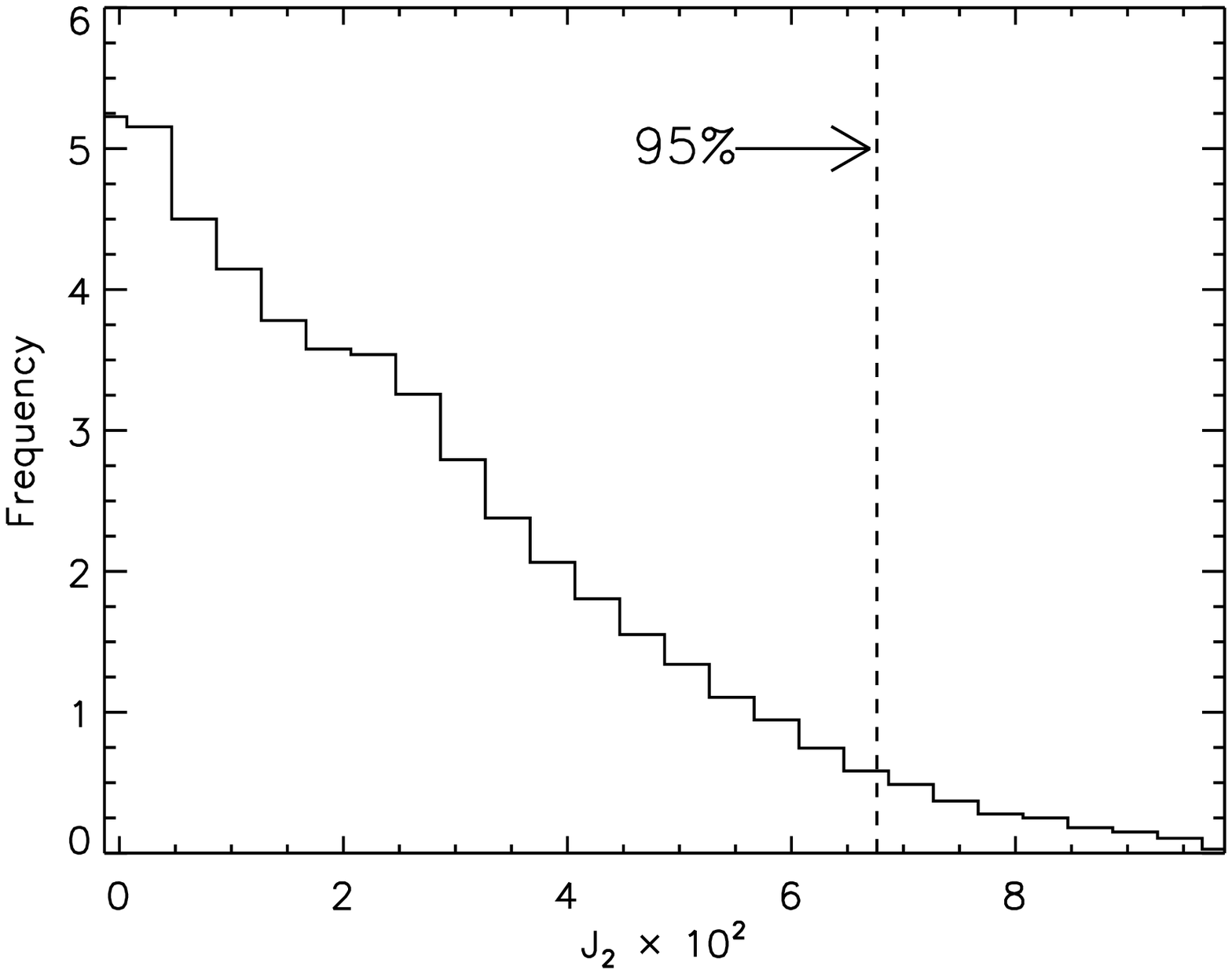}
\caption{ Posterior probability distribution for the quadrupole moment
  ($J_2$) of HD~189733b. See \S~\ref{sec:analysis} for a discussion of
  the underlying assumptions in this calculation.}
\label{fig:periodj2con}
\end{figure}

Conversely, if we assume HD~189733b to be synchronously rotating with
zero obliquity and $P_{\rm rot} = 2.218573$ days, then we may place an
upper bound on the rotationally-induced $J_2$, using
Eqn.~(\ref{eq:f}). The resulting posterior probability distribution
for $J_2$ is shown in Figure~\ref{fig:periodj2con}. We find that $J_2$
must be smaller than $0.068$ with $95\%$ confidence. To place this in
perspective, we note that if HD~189733b were a uniform-density sphere
then under the same assumption of spin-synchrony and zero obliquity
one would expect $J_2 = 0.0018$.\footnote{This was derived from
  Eqn.~(3) of Ragozzine \& Wolf~(2009), using a Love number $k_2=1.5$
  for a uniform-density sphere.} More centrally condensed planets
would lead to smaller values of $J_2$. Hence, the empirical upper
bound on $J_2$ is not constraining on physically plausible models of
the planet's interior.

\subsection{Uniform precession} \label{sec:conprec}

As discussed in \S~\ref{sec:prec}, if HD~189733b is significantly
oblate and oblique, then the spin axis will precess with a much
shorter period than the precession periods of Saturn or Jupiter. The
previous analysis assumed a fixed orientation for the planetary spin
axis, and therefore the results must be understood as valid only for
cases in which the precession period is much longer than 1~yr. The
actual precession period depends on the internal constitution of the
planet, specifically its moment of inertia and $J_2$. In the next part
of our analysis we relaxed the assumption of a fixed orientation for
the planetary spin axis, and allowed the axis to precess uniformly
about the orbital axis.

\begin{figure*}[htbp] %  figure placement: here, top, bottom, or page
   \centering
   \epsscale{1.0}
   \plotone{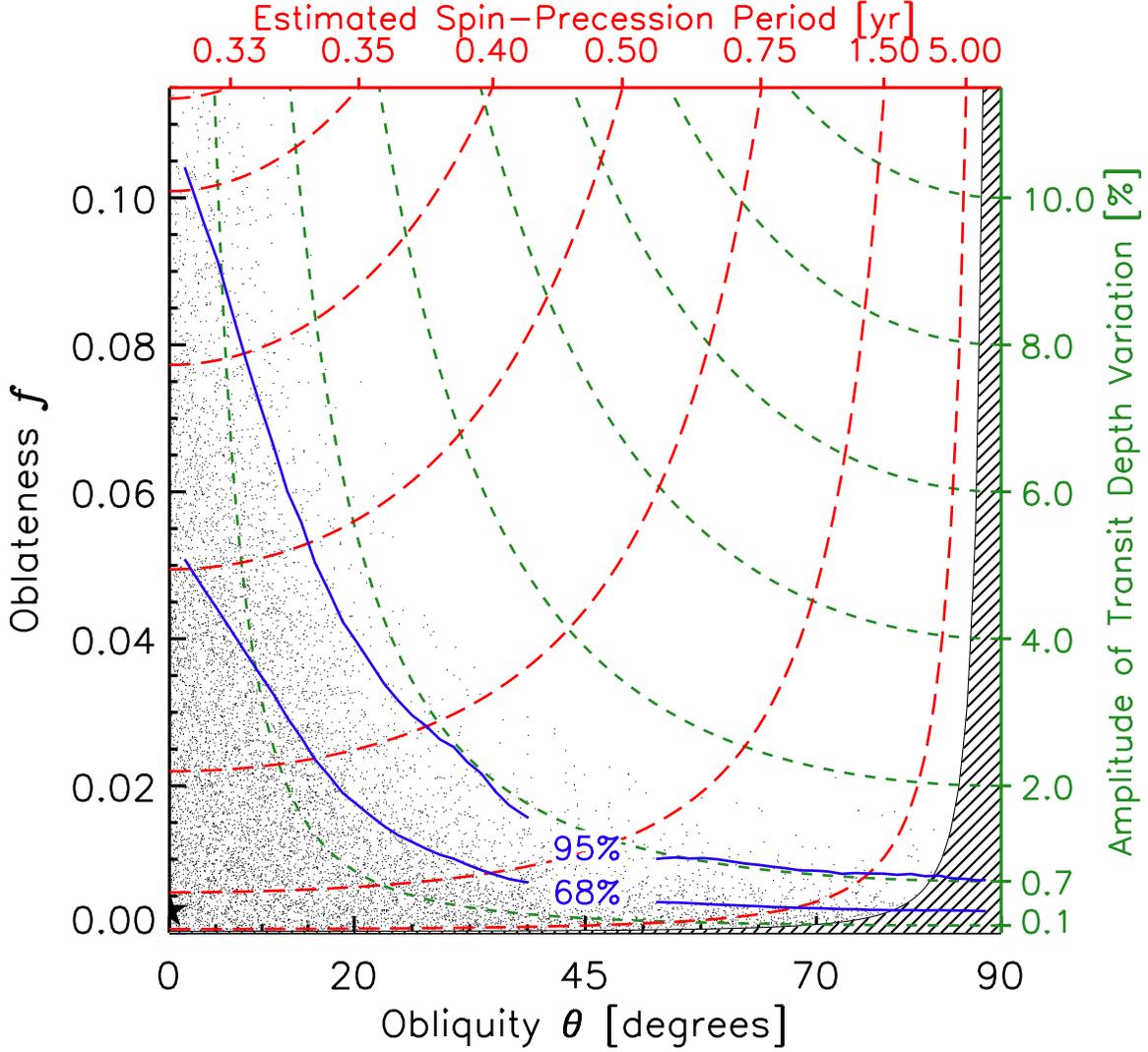}
   \caption{Constraints on the true oblateness and obliquity of
     HD~189733b, based on {\em Spitzer} observations of 7 transits.
     The solid blue curves bound the regions containing $68\%$ or
     $95\%$ of the MCMC samples in the oblateness-obliquity
     ($f$--$\theta$) plane. The black points are a random subsample of
     the full Markov chain, shown to illustrate the posterior
     probability density. The short-dashed green contours indicate the
     amplitude of transit depth variations. The long-dashed red
     contours show the expected spin-precession period, according to
     the formalism of \S~\ref{sec:prec}. The hatched region was not
     included in our analysis because the estimated precession period
     is $>$10~yr.}
   \label{fig:fiveyr}
\end{figure*}

We assumed that the {\it true} oblateness $f$ and {\it true} obliquity
$\theta$ are constant in time, and allowed the azimuthal angle $\phi$
to be a linear function of time,
\begin{eqnarray}
\phi(t) = \frac{2 \pi t}{P_{\rm prec}} + \phi_0,
\end{eqnarray}
for some precession period $P_{\rm prec}$ and initial phase angle
$\phi_0$. The time evolution of $\pf$ and $\po$ are then given by
Eqns.~(\ref{eq:projob}-\ref{eq:beta}). A consequence of this time
evolution is that the transit depth, $\delta(t)$, is variable. Apart
from small corrections due to limb darkening, the transit depth is
equal to the ratio of areas of the projected disks of the planet and
star:
\begin{eqnarray}
\delta(t) \approx \left(\frac{R_{\rm eq}}{R_\star}\right)^2 \left[ 1-\pf(t)\right].
\end{eqnarray}
For large oblateness and obliquity, the transit depth variations can
be easier to detect than the slight distortions in the ingress and
egress portions of a single light curve. For a Saturn-like oblateness
($f \approx 0.1$) at $45^\circ$ obliquity, the transit depth
variations are nearly $5\%$ (see Fig.~\ref{fig:fiveyr}). For the 7
{\it Spitzer} time series, the transit depths were found to agree with
one another to within 0.5\%. This suggests that if HD~189733b were as
oblate as Saturn and were also significantly oblique, then the
spin-precession period must be much longer than 1.6~yr. For a
quantitative and physically self-consistent analysis, we computed a
second MCMC, using the parameters $f$, $\theta$, $P_{\rm prec}$ and
$\phi_0$ to describe the shape of the planet,\footnote{It proved
  advantageous to use $R_{\rm eq} \sqrt{1-f}$ as a fitting parameter,
  rather than $R_{\rm eq}$, to reduce the correlation with $f$.}
rather than the sky projected parameters $\pf$ and $\po$.

In principle, we could have used Eqn.~(\ref{eq:prec}) to enforce a
relationship between $P_{\rm prec}$ and $f$, but that would have
required some assumptions about the interior density structure of the
planet. We preferred to allow $P_{\rm prec}$ to be a free parameter to
avoid making such assumptions. The structure of close-in planets may
differ from that of Jupiter and Saturn, due to a different history of
formation and evolution, as well as the tidal influence of the parent
star. We restricted our attention to $P_{\rm prec} < 10$~yr, because
for longer precession periods the orientation of the spin axis would
appear fixed over the 1.6~yr span of the {\it Spitzer} observations,
and therefore our analysis would revert to that of the previous
section.

We created two Markov chains, each with $10^6$ links, and merged them
after removing the first $10^5$ links of each chain.  The results are
depicted in Fig.~\ref{fig:fiveyr} as confidence regions in the
$f$--$\theta$ plane. This figure is analogous to
Fig.~(\ref{fig:constraintsallob}), although in this case we are
constraining the true oblateness and obliquity as opposed to the
projected quantities. Solutions with high oblateness and obliquity are
ruled out by the observed absence of transit depth variations.
Indeed, for $P_{\rm prec} < 10$~yr we can rule out a wider swathe of
obliquity-oblateness parameter space than we did under the assumption
of a nonprecessing planet.

For reference, Fig.~\ref{fig:fiveyr} also shows contours indicating
the amplitude of the expected transit depth variations, as well as
contours indicating the spin precession period calculated from
Eqn.~(\ref{eq:prec}), assuming $\mathds{C} = 0.225$. The precession
period is smaller than 10~yr over more than 90\% of the $f$--$\theta$
plane.

\section{Discussion} \label{sec:disc}

We have made the first attempt to measure the shape of a transiting
exoplanet. Using {\it Spitzer} observations of 7 transits of
HD~189733b, we have placed upper limits on the planet's oblateness.
The observed absence of lightcurve anomalies constrains the
sky-projected oblateness and obliquity at each epoch. The collection
of light curves spanning 1.6~yr constrains the true oblateness and
obliquity, because spin precession would have produced transit depth
variations in contradiction of the data. In both analyses an
oblateness as large as that of Saturn could be ruled out with $>$95\%
confidence.

The resulting upper limits on the oblateness are physically
meaningful, in the sense that they represent a physically possible
degree of oblateness comparable to that of giant planets in the Solar
system. However, for HD~189733b one would naturally expect a slower
rotation rate and a smaller oblateness than those of Saturn or
Jupiter, because tidal effects are expected to have slowed down the
planetary rotation until it was synchronized with the orbital
motion. Assuming that the planet is indeed spin-orbit synchronized
with zero obliquity, we were able to place an upper bound on the
quadrupole moment of the planet's density distribution: $J_2 <
6.8\times10^{-2}$ with 95\% confidence. However this is a weak upper
bound, in the sense that the theoretically expected value is at least
an order of magnitude smaller.

The planet HD~189733b was chosen for this analysis because of the
favorable stellar brightness and transit depth, as well as the large
collection of high-quality data that are currently available for this
system. However, this system suffers from the drawback that the
theoretically expected oblateness is smaller than detection
thresholds. The transiting planets HD~17156b (Barbieri et al.~2007)
and HD~80606b (Moutou et al.~2009, Fossey et al.~2009, Garcia-Melendo
\& McCullough~2009) are attractive targets because they have much
longer periods and higher orbital eccentricities than HD~189733b;
although tidal evolution is expected to be important in both of those
cases, it may have resulted in ``pseudosynchronization'' (Hut 1981) or
some other state besides spin-orbit synchronization.

Furthermore, the {\it Kepler} mission\footnote{{\tt
    http://kepler.nasa.gov/}; see also Borucki et al.~(2009).} and
other surveys for transiting planets should soon find planets with
longer orbital periods and larger orbital distances, for which there
is no reason to expect spin-orbit synchronization. There should be a
``sweet spot'' in orbital distance, far enough that the planet might
be expected to rotate as rapidly as Jupiter or Saturn, yet close
enough for the stellar gravitational field to cause relatively rapid
spin precession that would be manifested as transit depth variations.
A more general analysis of precession-induced transit depth variations
seems warranted.

\acknowledgements We are very grateful to Heather Knutson, Eric Agol,
and their collaborators, for organizing the {\it Spitzer}
observations. We are indebted to Dan Fabrycky for pointing out that
spin precession can play an important role in the analysis. We thank
Darin Ragozzine and the referee, Jason Barnes, for providing timely
and detailed comments on the manuscript. We also benefited from
discussions with Sara Seager and Saul Rappaport.

\begin{appendix}
\section{A numerical routine for fast, efficient calculation\\ of transit light curves for oblate exoplanets} \label{app:algorithm}

In this appendix we describe a method for fast and stable calculations
of light curves of ellipsoidal exoplanets, for which the sky
projection is an ellipse. The development of this algorithm was
crucial for our analysis, to make the MCMC method computationally
tractable. Proper convergence of these posterior distributions
required many millions of executions of the light curve
calculation. Seager \& Hui (2002) and Barnes \& Fortney (2003) both
calculated light curves of oblate planets, but for those authors the
computation time was probably of secondary importance.

\begin{figure}[htbp] %  figure placement: here, top, bottom, or page
   \centering
   \epsscale{0.7}
   \plotone{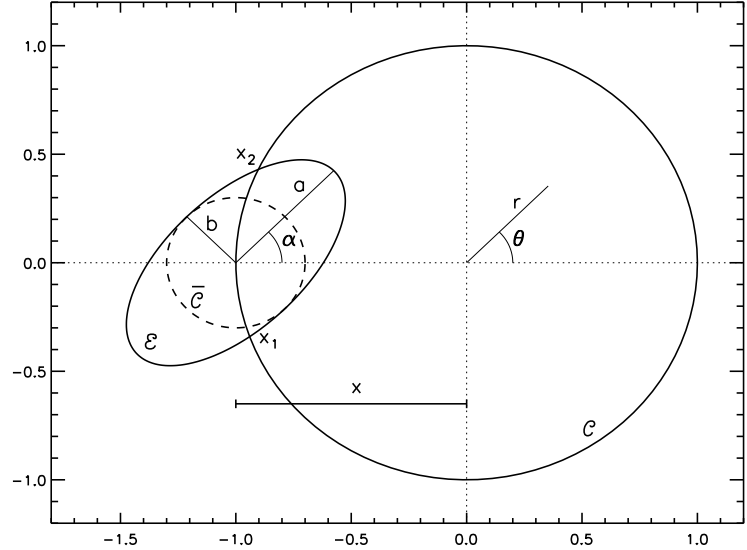} 
   \caption[Geometrical configuration for the transit of an
   ellipsoidal planet across a spherical star.]{{ Geometry of the
       transit of an ellipsoidal planet across a circular stellar
       disk.}  The deficit in flux due to the obscuring ellipse with
     semi-major axis $a$, semi-minor axis $b$ and orientation $\alpha$
     a distance $x$ from the center of the star may be found by
     integrating the brightness profile of the star over the
     intersection region of the ellipse ${\cal E}$ and circle ${\cal
       C}$. The ellipse and circle intersect at points $\vec{x}_{1,2}$.  A
     closed-form solution for the flux deficit exists for the
     inscribed circle $\bar{\cal C}$ of radius $b$. Only the
     intersection region less this circle, ${\cal E}/\bar{\cal C} \cap {\cal
       C}$ needs to be integrated numerically.}
   \label{fig:setup}
\end{figure}

Figure~\ref{fig:setup} illustrates the basic geometry of the
calculation. An elliptical shadow with major and minor axes $a$ and
$b$ obscures a portion of the circular stellar disk.  The
projected-oblateness $\pf$ is defined as $(a-b)/a$. Distances are
expressed in units of the stellar radius, $R_\star$. The semimajor
axis of the ellipse is inclined by an angle $\alpha$ from the line
connecting the centers of the ellipse and circle. The angle $\alpha$
is generally not the same as $\po$, but the angles are equal for a
central transit. The centers of the star and planet are separated by a
distance $x$. We assume the stellar disk has a brightness profile
$I_\star(r,\theta)$ including, for example, a radial limb-darkening
profile. The fractional flux deficit, $F(x; a,b, \alpha, I_\star)$,
due to the obscuring ellipse is
\begin{eqnarray}
F(x; a,b, \alpha, I_\star) = \frac{1}{F_0} \int_{{\cal E} \cap {\cal C}}
I_\star(r,\theta) ~r ~dr~d\theta \label{eq:ff}
\end{eqnarray}
where $F_0$ is the total unobscured flux and the integral is performed
over the region bounded by the intersection of the ellipse and circle,
denoted as ${\cal E} \cap {\cal C}$.  We consider a radial brightness
profile given by a quadratic limb-darkening law,
\begin{eqnarray}
	I_\star(r,\theta)& = &I_\star(r; u_1, u_2) \nonumber \\
		&  = & I_\star(1) \left[1- u_1\left(1-\sqrt{1-r^2}\right)-u_2 \left( 1-\sqrt{1-r^2}\right)^2 \right] \label{eq:ld}
\end{eqnarray}
where $u_1$ and $u_2$ are the limb darkening parameters. With this
law, $F_0/I_\star(1) = \pi (1- 1/3 u_1-1/6 u_2)$. We denote the
quadratic-profile fractional flux deficit by the form $F(x; a, b,
\alpha, u_1, u_2)$. For $a = b = R_p/R_\star$, $F(x; a, u_1, u_2)
\equiv F(x; a, a, \alpha, u_1, u_2)$ has a closed-form analytic
solution in terms of elliptic integrals (Mandel \& Agol 2002). The
problem is more complicated for arbitrary $a$, $b$, $\alpha$, $u_1$
and $u_2$ since the vertices of the intersection region ${\cal E} \cap
{\cal C}$ cannot always be determined analytically, and the integral
cannot be done in closed form.

There are standard routines for computing the intersections of two
ellipses that would allow us to define the intersection region ${\cal
  E} \cap {\cal C}$ (see, e.g., Hill 1994).  We found, however, that
these algorithms are not numerically stable over the entire region of
parameter space of interest. We chose to sacrifice runtime to ensure
stability by finding the intersections of a polygonal representation
of the obscuring ellipse and the disk, as follows. First, we
analytically find the coordinates in the $x-y$ plane of points on the
boundary of the ellipse at $N$ uniformly selected angles (from $0$ to
$2 \pi$).  Next, these points in the plane are connected by line
segments.  Finally, the routine accumulates approximate intersections
of the ellipse and circle by determining the intersections of each
line segment and the circle. We use $N \approx 200$ to ensure adequate
precision.

With the points of intersection determined, we move on to the
calculation of the integral in Eqn.~(\ref{eq:ff}). Before discussing
the general problem, we first consider the case of uniform brightness
($u_1=u_2 =0$) and solve for $F(x; a,b, \alpha, 0, 0)$.  Here,
solving the integral in Eqn.~(\ref{eq:ff}) is equivalent to finding
the area of the region ${\cal E} \cap {\cal C}$. This area may be
calculated analytically by using the formula for the area of an
elliptical chord $A(\theta_1,\theta_2, a, b)$:
\begin{eqnarray}
	A(\theta_1,\theta_2,a,b) & = & a b \left[ (\theta_1-\theta_2) - \sin(\theta_1-\theta_2)\right]/2
\end{eqnarray}
where the angles $\theta_{1,2}$ are measured relative to the
semi-major axis.  To solve for $F(x; a,b, \alpha, 0, 0)$ we add the
area of the elliptical chord (defined by the lines connecting the
intersection points\footnote{In general, a circle and an ellipse may
  intersect at up to four distinct locations; for an exoplanet with
  small oblateness and $R_{\rm p} \approx R_{\rm Jupiter}$ orbiting a
  star with $R_\star \approx R_\sun$, only two intersections (at most)
  are expected. } and the curve bounding the ellipse that is internal
to the stellar disk) to the area of the circular chord (defined by the
complement of this elliptical chord and ${\cal E} \cap {\cal C}$).

Next, consider the integral given by Eqn.~(\ref{eq:ff}) for nonzero
$u_1$ and $u_2$.  Numerical integration must be performed.  We could
perform two-dimensional numerical integration over ${\cal E} \cap
{\cal C}$ [as was done by Seager \& Hui (2002)] or numerically
integrate in the radial direction, calculating the intersections of
circle and ellipse at each integration step [as was done by Barnes \&
Fortney (2003)]. For our application we chose the former, although
because of the difficulty in specifying the integration region, we
found standard deterministic integration techniques to be too slow.

Instead, we chose to use Monte Carlo integration [see, e.g., Press et
al.\ (2007)]. With Monte Carlo integration, the integral $\int_{\cal
  A} f(x,y)~dx~dy$ over the region ${\cal A}$ is calculated by
sampling the function $f(x,y)$ at $N$ uniformly distributed random
points in a region ${\cal R}$ that covers ${\cal A}$. When a sample
$(x,y)$ is drawn from ${\cal R}$ that is not in ${\cal A}$, we set
$f(x,y) = 0$. According to the fundamental theorem of Monte Carlo
integration,
\begin{eqnarray}
\int_{\cal A} f(x,y)~dx~dy &  \approx & {\rm Area}[{\cal R}]
\left[\langle f \rangle \pm \sqrt{\frac{{\rm Var}(f)}{N}}\right].
\end{eqnarray}
The means are calculated over the $N$ randomly sampled points (Press
et al.\ 2007).  In general, Monte Carlo integration is reserved for
integrals of high dimensionality due to its slow convergence
($1/\sqrt{N}$) compared to more traditional methods when calculating
two-dimensional integrals. For our purposes, Monte Carlo integration
reduces the difficulty associated with bounding the intersection
region ${\cal E} \cap {\cal C}$. An improvement in convergence ($\sim
1/N$) is obtained by sampling ${\cal R}$ ``quasi-randomly'' as opposed
to randomly, using a low-discrepancy random sequence called the Sobol'
sequence (Sobol' \& Shukhman 1995, Press et al.\ 2007). The Sobol'
sequence is a sequence of uniformly distributed values on the unit
interval, $s_i$ for $i \ge 0$, such that a given sequence member at
index $I$ is {\em maximally} distant from all previous samples $i <
I$.

To simplify the computational effort required for the integral of
Eqn.~(\ref{eq:ff}) when using the quadratic limb-darkened profile
[Eqn.~(\ref{eq:ld})], we rewrite the integral as
\begin{eqnarray}
  F(x; a, b, \alpha, u_1, u_2) & \times& \pi \left( 1- \frac{u_1}{3}-\frac{u_2}{6} \right)\nonumber \\
		   & = & \int_{{\cal E} \cap {\cal C}} 1- u_1\left(1-\sqrt{1-r^2}\right)-u_2 \left( 1-\sqrt{1-r^2}\right)^2 ~r ~dr~d\theta \nonumber\\
		   & = & \left(1-u_1-u_2\right)\left(\int_{{\cal E} \cap {\cal C}} ~r~dr~d\theta\right)\nonumber\\
		   &&\;\;\;\;\;\;\;\;+ \int_{{\cal E} \cap {\cal C}} \left[\left(u_1+2 u_2\right) \sqrt{1-r^2}-u_2 \left(1-r^2\right)\right]~r~dr~d\theta \nonumber\\
		   & = & \left(1-u_1-u_2\right) \times F(x;a,b, \alpha, 0, 0)\nonumber\\
		   &&\;\;\;\;\;\;\;\;+ \int_{{\cal E} \cap {\cal C}} \hat{I}_\star(r; u_1,u_2)~r~dr~d\theta \label{eq:splitld}
\end{eqnarray}
where $F(x;a,b, \alpha, 0, 0)$ is the uniform-brightness solution
(calculated as described above) and $\hat{I}_\star(r; u_1,u_2) =
\left(u_1+2 u_2\right) \sqrt{1-r^2}-u_2 \left(1-r^2\right)$ describes
the nonuniform component of the brightening profile.
$\hat{I}_\star(r = 1; u_1,u_2) = 0$ at the stellar limb.  The remaining unknown integral is over a
function that has low variance over any bounding region ${\cal R}$
covering the limb, making the absolute error in the Monte Carlo
integration smaller (for a fixed $N$) as compared to integrating
$I_\star$ in full.

Note that the only portion that requires numerical integration is the
intersection region excluding the circle $\bar{\cal C}$ of radius $b$
inscribed in ${\cal E}$ [see Fig.~(\ref{fig:setup})]. We denote this
smaller region by ${\cal E}/ \bar{\cal C} \cap {\cal C}$.  The
integral over $\bar{\cal C}$ is analytic, as per Mandel \&
Agol~(2002), and is denoted $F(x; b, u_1, u_2)$. Thus
Eqn.~(\ref{eq:splitld}) becomes
\begin{eqnarray}
	F(x; a, b, \alpha, u_1, u_2) &\times& \pi\left( 1- \frac{u_1}{3}-\frac{u_2}{6} \right) \nonumber \\
			& = & \left(1-u_1-u_2\right) \times \left[F(x;a,b, \alpha, 0, 0)-F(x;b, 0,0)\right]\nonumber\\
			&&\;\;\;\;\;\;\;\;+ F(x;b, u_1,u_2) \nonumber \\
			&&\;\;\;\;\;\;\;\;+ \int_{{\cal E} \cap {\cal C}/\bar{\cal C}} \hat{I}_\star(r; u_1,u_2)~r~dr~d\theta. \label{eq:ffinal}
\end{eqnarray}
The only nontrivial component is the final term, which we denote
${\cal I}$ and for which we use quasi-Monte Carlo integration. For
slightly oblate exoplanets or weakly limb-darkened brightness
profiles, the contribution of this integral in the total flux deficit
is small compared to the remaining terms in Eqn.~(\ref{eq:ffinal}).
The absolute contribution of this term to the flux deficit is bounded
by
\begin{eqnarray}
\frac{1}{\pi} \left( 1- \frac{u_1}{3}-\frac{u_2}{6} \right)^{-1} {\cal I}
& \le &
\frac{ u_1+2 u_2}{1-  1/3 u_1- 1/6 u_2} \times \left(a b- b^2\right) \nonumber  \\
& = & a b ~\pf \frac{ u_1+2 u_2}{1-  1/3 u_1- 1/6 u_2}. 
\end{eqnarray}
For comparison, the size of the flux deficit in the absence of
limb-darkening is $a b$.
	
\begin{figure}[htbp] %  figure placement: here, top, bottom, or page
   \centering
   \epsscale{1.0}
   \plotone{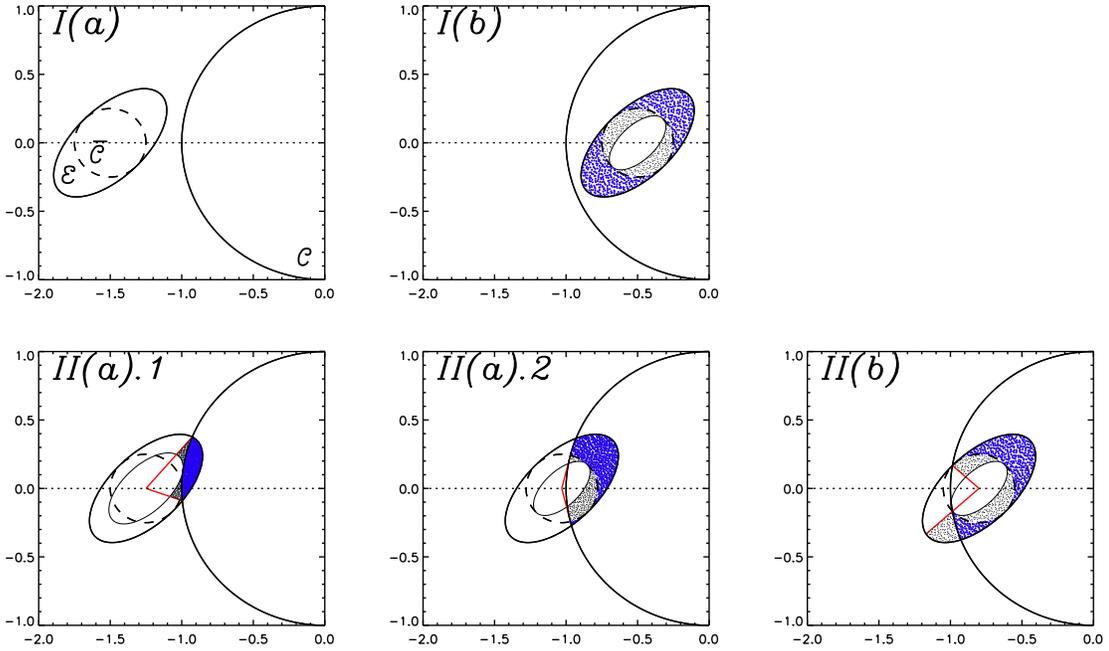} 
   \caption[Quasi-Monte Carlo integration of the nontrivial component
   of the total flux deficit for the stellar transit of an oblate
   planet.]{{ Quasi-Monte Carlo integration of the nontrivial
       component of the total flux deficit for the stellar transit of
       an oblate planet.}  These figures show the transit phases and
     quasi-Monte Carlo integration regions described in
     Appendix~\ref{app:algorithm}, which are needed to evaluate the
     nontrivial integral ${\cal I}$ [see Eqn.~(\ref{eq:ffinal})].  The
     labels in the upper left hand corners of each figure correspond
     to those in the text.  In each figure, the inner ellipse with
     semi-major axis $a_1$ gives the inner boundary of the elliptical
     annular sector.  The red rays emanating from the ellipse indicate
     the angular extent of the sector (with angles $\theta_{1,2}$
     relative to the semi-major axis).  The blue or black points are
     $1000$ uniformly distributed Sobol' points in the elliptical
     annular sector.  The blue points are those which fall in the
     integration region ${\cal E} /\bar{\cal C}\cap {\cal C}$. }
   \label{fig:Iint}
\end{figure}

To calculate ${\cal I}$ via quasi-Monte Carlo integration, we select a
covering region, ${\cal R} \supset {{\cal E}/\bar{\cal C} \cap {\cal
    C}}$, that efficiently bounds the integration region and that can
be easily sampled with the Monte Carlo technique (${{\cal E}/\bar{\cal
    C} \cap {\cal C}}$ cannot). Here, easily sampled means that we may
take a two dimensional Sobol' sequence, uniform in $[0,1]\times[0,1]$,
and analytically transform it such that the transformed sequence
uniformly samples our chosen covering region. A uniform sampling on
the unit square may be mapped to a uniform sampling over a region
bounded by an elliptical annular sector (see
Appendix~\ref{ap:sampling} for details).  An elliptical annular sector
is the region bounded between two concentric ellipses (i.e., two
ellipses with coincident centers and equal axis ratios and position
angles) with semi-major axes $a_1 < a_2$ and by the rays (emanating
from the common center) at angles $\theta_1$, $\theta_2$ relative to
the semi-major axis.  The formulae for the specific values of
$\theta_1$, $\theta_2$, $a_1$ and $a_2$ as functions of the parameters
$x$, $a$, $b$, and $\alpha$ for use in the integration routine are
chosen from one of the following cases [with reference to
Fig.~(\ref{fig:Iint})]:
\begin{tabbing}
{\em (I)} \= {\em One or fewer intersections between circle and ellipse}.    \\
\> {\em (a)} \= $ x > 1$; The obscuring ellipse is external to the circle, ${\cal I}$ = 0. \\
\> {\em (b)} \=$ x < 1$; The obscuring ellipse is properly contained in the circle; \\
\>\>$\theta_1 = 0$, $\theta_2 = 2 \pi$, $a_1 = b$, $a_2 = a$. \\
{\em (II)} {\em Two points of intersection between circle and ellipse.} \\
\>The circle and ellipse intersect at $x$-coordinates $x_{1,2}$ at angles $\theta_{1,2}'$ from the $x$-axis. \\
\> {\em (a)} $x > 1$; For each intersection $x$-coordinate $x' = x_{i}$, angle $\theta' = \theta_i'$ \\
\>\>and final solution angle $\theta = \theta_i$: \\
\> \>{\em (1)} \= $ |x| > |1/x'| $; The line connecting the point of intersection and the ellipse \\
\>\>\>center [at (x,0)] intersects the circle exactly once: $\theta = \theta'-\alpha$. \\
\>\> {\em (2)} \= $ |x| < |1/x'|$;  The line connecting the point of intersection and the ellipse center \\
\> \>\> [at (x,0)] intersects the circle  twice: use $\theta = \hat{\theta}-\alpha$ where the angle \\
\>\>\> $\hat{\theta} =  \tan^{-1} 1/\sqrt{1-x^2}$ defines the line segment that connects \\
\>\>\> the ellipse center and the point of tangency on the circle.\\
\>\> If $a'$ is the semi-major axis of the concentric ellipse ``kissing'' the circle\footnotemark~ then\\
\>\> $a_1 = {\rm max}(b, a')$, $a_2 = a$. \\
\> {\em (b)} \=$ x < 1$; Use the angle $\theta_{b}$ defined by the intersection points of the inscribed circle  \\
\> \>with radius $b$ ($\bar{\cal C}$) and the stellar disk (${\cal C}$): $\theta_{1,2} = \pm \theta_b-\alpha$, $a_1 = b$, $a_2 = a$.
\end{tabbing}
\footnotetext{``Kissing" ellipses intersect at exactly one point.  The ``kissing'' ellipse's semi-major axis $a'$ for this application is found with a linear search.} 
By choosing the elliptical annular sector as defined above as the
bounding region for the the desired integration region $\left({{\cal
      E}/\bar{\cal C} \cap {\cal C}}\right)$, we ensure that the
integration region covers approximately $50\%$ of the bounding region
regardless of the values of $x$, $a$, $b$, and $\alpha$.

To give some idea of the computation time, we consider the case $x =
1$, $a = 0.155$, $b = 0.148$, $u_1 = 0.2$, $u_2 = 0.3$, and $\alpha =
0.5$, corresponding to an oblate ($\pf = 0.05$) version of HD~189733b.
Using a C++ implementation of our algorithm on a $2.6$ GHz Intel Core
2 Duo MacBook Pro, it takes 0.5~ms to compute $F$ with a precision of
1~ppm. For comparison, it takes 0.004 ms to execute the same
computation for a spherical planet, using the Fortran implementation
of the code by Mandel \& Agol (2002).

\section{Uniform sampling of an elliptical annular sector} \label{ap:sampling}

\subsection{Elliptical annular sector}

\begin{figure}[htbp] %  figure placement: here, top, bottom, or page
   \centering
   \epsscale{0.5}
   \plotone{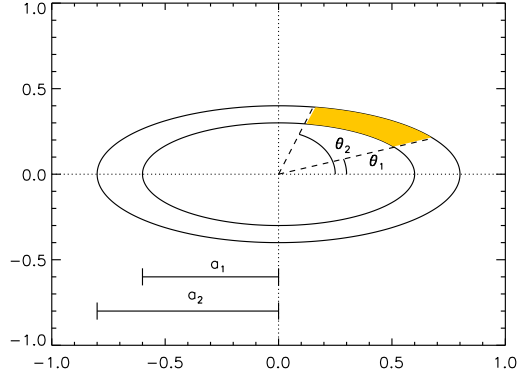} 
   \caption{{  An elliptical annular sector.}}
   \label{fig:eas}
\end{figure}

A point $(x,y)$ is inside the elliptical annular sector centered at
$(0,0)$ with semi-major axis in the $x$-direction, axis ratio
$\epsilon$, inner radius $a_1$, outer radius $a_2$, and sector angles
$\theta_{1,2}$ if the all of following conditions are satisfied:
\begin{eqnarray}
	(1) &\;\;& \frac{x^2}{a_1^2}+\frac{y^2}{\left(\epsilon a_1\right)^2} > 1\\
	(2) &\;\;& \frac{x^2}{a_2^2}+\frac{y^2}{\left(\epsilon a_2\right)^2} < 1 \\
	(3) &\;\;& \mbox{The line connecting $(0,0)$ to $(x,y)$ is at an angle $\theta$ relative to the $x$-axis such that $\theta_2 > \theta > \theta_1$ }
\end{eqnarray}
An elliptical annular sector is illustrated in Fig.~(\ref{fig:eas}).

\subsection{Uniform sampling}
Let $(u, v)$ be a uniform sample of the unit square $[0,1]
\times[0,1]$.  Then $(u',v')$ is a uniform sample of the elliptical
annular sector where
\begin{eqnarray}
	u' & = & a_2 r \cos(\theta) \\
	v' & = & \frac{a_2 r}{\epsilon} \sin(\theta)
\end{eqnarray}
and
\begin{eqnarray}
	r & = & \sqrt{(1-u) ~a_1^2+u} \\
	\theta & = & (1-v) \tan^{-1} (\epsilon \tan \theta_1)+v  \tan^{-1} (\epsilon \tan \theta_2).
\end{eqnarray}

\end{appendix}

\end{document}